\documentclass{aa} 
\usepackage{float} 
\usepackage{graphicx}  
\usepackage{graphics}
\usepackage{txfonts}  
\usepackage{natbib}
\usepackage{adjustbox}
\usepackage{booktabs}
\usepackage{mathtools}
\usepackage{comment}
\bibpunct{(}{)}{;}{a}{}{,} 
\usepackage{multicol}
\usepackage{placeins}
\usepackage{longtable}
\usepackage[dvipsnames]{xcolor}
\usepackage[normalem]{ulem}
\usepackage{bm}
\usepackage{braket}
\usepackage{tabularx}
\usepackage{hyperref}
\hypersetup{
    colorlinks=true,
    linkcolor=blue,
    citecolor=blue,
    filecolor=magenta,      
    urlcolor=cyan,
    pdftitle={Overleaf Example},
    pdfpagemode=FullScreen,
    }

\begin{document}

\title {Dilution of accreted planetary matter in hot DA white dwarfs according to their mass}
\subtitle{}

\author{M. Deal\inst{1} \and S. Vauclair\inst{2} \and S. Charpinet\inst{2} \and G. Vauclair\inst{2}}
  
\institute{LUPM, CNRS, Universit\'e de Montpellier, Place Eug\`ene Bataillon, 34095 Montpellier, France\\
\email{morgan.deal@umontpellier.fr} 
           \and
IRAP, CNRS, Universit\'e de Toulouse, CNES, 14 Avenue Edouard Belin, 31400 Toulouse,
France
            }
           
\date{\today}
\abstract
{A large proportion of observed white dwarfs show evidence of debris disks, remnants of the former planetary systems, and/or signatures of heavy elements in their atmospheres, induced by the accretion of planetary matter onto their surfaces. The observed abundances are the result of the balance between the accretion flux and the dilution of this planetary material by internal transport processes.  A recent study showed that more massive DA white dwarfs are less polluted than smaller mass ones. It was suggested that the reason could be related to the formation of planetary systems when these stars were on the main sequence.} 
{The aim of this work is to test how internal dilution processes, including thermohaline convection, change with white dwarf masses, and whether such an effect could account for  variations in the observed pollution.}
{We computed the efficiency of atomic diffusion and thermohaline convection after the accretion of heavy elements onto white dwarfs using static DA models with various masses, effective temperatures, and hydrogen contents.}
{We confirm that thermohaline convection is always more efficient in diluting accreted elements than atomic diffusion, as previously shown in the literature. However, we find that element dilution by thermohaline convection is less efficient in massive white dwarfs than in smaller mass ones, due to their larger internal density.}
{We showed that the differences in observed heavy element pollution in white dwarfs according to their masses cannot be explained by the dilution induced by atomic diffusion and thermohaline mixing alone. Indeed, the pollution by planetary system accretion should be more easily detectable in massive white dwarfs than in low-mass ones. We discuss other processes that should be taken into account before drawing any conclusion about the occurrences of planetary systems according to the mass of the star on the main sequence.}

  
\titlerunning{Dilution of accreted planetary matter in hot DA white dwarfs according to their mass}
  

\maketitle 
\nolinenumbers
\section{Introduction}

All stars with masses lower than $8$M$_\odot$ become white dwarfs (WDs) at the end of their evolution. They represent about $97\%$ of the stellar population of the Galaxy. A key question is how planetary systems are affected when their host stars evolve from the main sequence up to the final WD stage. Recent observations give evidence that many of these planetary systems have partly survived, leading to debris disks around the WDs detected in the infrared \cite[see][and references therein]{zuckerman87,farihi08,guidry24} or by their transit in front of the star \cite[see][and references therein]{vanderburg15,vanderbosch20,vanderbosch21,bhattacharjee25}. Planetary matter falling onto WD atmospheres is also detected as heavy-element lines in their spectra \citep[see for example][and references therein]{zuckerman10,gentille-fusillo21}.

The accreted elements cannot remain in the WD atmospheres because of rapid downward diffusion. The observed abundances are therefore the result of a delicate balance between the accretion rate and the downward dilution rate. The situation is still more complicated due to the fact that, in addition to atomic diffusion, the accumulation of heavy matter above lighter matter can give rise to thermohaline (fingering) mixing, a well-known process in the oceans that also occurs in stars \citep[][and references therein]{deal13,bauer18,bauer19,wachlin22}.

In most published studies, however, atomic diffusion is the only process introduced for metal settling below the atmospheres of WDs. In the present work, we point out the necessity of a correct treatment of element dilution inside the stars to derive reliable results on the chemical composition of the planetary matter that remains around WDs. As mentioned above, thermohaline mixing may have a much larger diluting effect than atomic diffusion, in which case the results obtained with atomic diffusion alone are not correct. It is even more important considering recent attempts to derive constraints on the pristine planetary systems and the circumstances of their formation, as discussed below.

\cite{harrison21} analysed 230 helium-rich, cool WDs, with metal-polluted atmospheres, particularly in Ca, Mg, Fe, Ti, Ni, Cr, and Na. They performed direct modelling of the expected chemical composition of the WD atmospheres, assuming many different initial compositions, temperatures, and geological histories of the original bodies as they were formed in the protoplanetary system. Then they used a Bayesian approach to derive the most likely original abundances of the accreted material from the observed abundances in their sample of WDs. They were able to infer the chemical composition of the original planetary bodies as well as physical processes such as heating and fragmentation during their formation. However, in their analysis, thermohaline mixing was ignored.

\cite{ouldrouis24} (hereafter OR24) performed a statistical study of the metal abundances observed in more than 250 hot WDs (13~000~K < $T_\mathrm{eff}$ < 30~000~K) according to their masses. They found that more than $40\%$ of all these WDs show silicon and sometimes carbon lines, presumably due to planetary matter accretion. Surprisingly, they also found evidence of a metallic trend according to the WD masses for similar effective temperatures. The fraction of polluted stars is around $44\%$ for the less massive ones (<$0.7$M$_\odot$), whereas it is only around $11\%$ for the more massive ones (>$0.8$M$_\odot$). Going backwards along evolutionary tracks, they claim that the most massive WDs are the descendant of main-sequence stars more massive than $3.5$M$_\odot$, and that their results give evidence of a difference in planetary formation at the origin, according to the stellar mass. This is an important result that needs to be confirmed. Here again, thermohaline mixing was ignored in the computations. More recently, \cite{cunningham25} revisited this question by computing population synthesis of WDs according to their masses. They proposed two possible scenarios to explain the observed trend, with no thermohaline convection.

Thermohaline convection was discussed in a simplified way, together with other hydrodynamical processes that may influence the abundances in polluted WDs, by \cite{buchan25}. In this important paper, they confirm that hydrogen-rich WDs may suffer enough thermohaline convection to inhibit the effect of atomic diffusion.

\cite{rogers25} analysed the surface composition of eight hydrogen- and helium-rich WDs, compared it with the composition of their disk, and found a good correlation in silicate mineralogy. This is also an indication that the diluting processes are not strongly selective for Mg and Si, which is consistent with a dilution by atomic diffusion (mostly dominated by the effect of gravitational settling) and thermohaline mixing.

As was first discussed by \cite{deal13} and confirmed in later papers \citep[see][and references therein]{wachlin22}; thermohaline mixing is much less efficient in He-rich (DB) WDs than in H-rich (DA) ones (see below). This is due to the initially heavier material, the larger gas density, and the larger convective depth in DBs compared to that of DAs. In the case of \cite{harrison21}, since they used a sample of solely DB WDs, we
have checked that their results are not affected by thermohaline mixing and are thus reliable. In the present paper, we focus on the study by OR24 that was carried out for a sample of DA WDs.

The paper is structured as follows: we describe the internal structure models of WDs and the modelling of accretion and transport processes in Sec.~2. The comparison of the efficiency of atomic diffusion and thermohaline convection for the different models and cases are presented in Sec.~3. The results are discussed in Sec.~4, and we conclude in Sec.~5.
\section{Modelling of the transport of chemicals after accretion}\label{sect:mod}

\subsection{The white dwarf structures}

\begin{table}
\caption{Properties of the white dwarfs' static structures.
}
\label{tab:mod}
\begin{tabular}{lcccc}     
\hline                     
Model & $M$ [M$_\odot$] & $R$ [R$_\oplus$] & $T_\mathrm{eff}$ [$10^3$~K] & $M_H/M_\mathrm{WD}$ \\
\hline
M06T20H-4 & $0.6$ & $1.48$ & $20$ & $10^{-4}$  \\
M08T20H-4 & $0.8$ & $1.16$ & $20$ & $10^{-4}$  \\
M10T20H-4 & $1.0$ & $0.90$ & $20$ & $10^{-4}$  \\
M06T25H-4 & $0.6$ & $1.53$ & $25$ & $10^{-4}$  \\
M08T25H-4 & $0.8$ & $1.18$ & $25$ & $10^{-4}$  \\
M10T25H-4 & $1.0$ & $0.91$ & $25$ & $10^{-4}$  \\
M06T30H-4 & $0.6$ & $1.57$ & $30$ & $10^{-4}$  \\
M08T30H-4 & $0.8$ & $1.20$ & $30$ & $10^{-4}$  \\
M10T30H-4 & $1.0$ & $0.92$ & $30$ & $10^{-4}$  \\
\hline
\end{tabular}
\end{table}

We want to compare the relative efficiencies of atomic diffusion and thermohaline convection in WD models corresponding to those discussed in OR24. The aim was to test in which cases atomic diffusion alone may be used to derive the accreted abundances from the ones observed in the stellar atmospheres, and in which cases thermohaline convection cannot be neglected. WD structures were computed using the STELlar modelling code of the Université de Montréal \citep[STELUM;][]{bedard22}. We relied on parametrised static models of DA WDs described in \cite{giammichele16,giammichele17}. Such models are typically used for WD asteroseismology, but their flexibility to explore the model parameter space makes them practical tools for our present purposes as well. The static models that we computed assume a core composition as predicted by evolution calculations \citep[from the BASTI code;][]{salaris10}, which is therefore mass-dependent. The helium-dominated mantle is a mixed C/O/He buffer of ~$10^{-1.3}$~M$_\odot$, whose structure is fixed for the computed set of models. It is surrounded by a pure hydrogen envelope, the mass of which may be modified. Once the chemical stratification of the DA WD was specified with its global parameters, the stellar structure equations were solved to establish the equilibrium state of the star. The equation of state, opacities, and convection treatment used in this process are those described in \cite{bedard22}. The depths of the outer hydrogen zones were estimated after seismic determinations \citep[see e.g.][]{romero25}. The list of models is presented in Table~\ref{tab:mod}.

\subsection {The accretion process}\label{sect:accr}

The precise way in which the matter of debris disks is accreted onto WDs is unclear. It is possible that accretion first occurs in stellar equatorial regions and that the accreted matter then spherically diffuses in the outer stellar layers. In our computations, we assumed as a first boundary condition that the matter is initially dispersed in a spherical shell at the surface. When stars have an outer convective zone, the depth of this shell is that of the convective layers. This is not the case for the effective temperature range studied here.

We were thus obliged to arbitrarily define the depth of this initial outer dilution zone. If all the mass was accreted in the first layer of the model (corresponding to an exterior mass ranging between $1.4\times10^{11}$ and $3.3\times10^{12}$~g, and a depth ranging between $0.3$ and $2.6$~km, depending on the model), the mean molecular weight in this layer would be very large compared to the composition of the star, leading to almost instantaneous thermohaline mixing. The situation would evolve and lead to a fully mixed region at the top of the star and an abundance of heavy elements decreasing with depth. To simulate this effect, we arbitrarily considered a layer with a mass $M_\mathrm{mix}=10^{15}$~g (corresponding to a depth between $39$ and $141$~km depending on the model), and we assumed additional dilution of the accreted matter that decays exponentially with the radius. The mass fraction profile of the accreted matter $X_\mathrm{accr}$ is defined as
\begin{equation}
   X_\mathrm{accr}=\frac{M_\mathrm{accr,mix}}{M_\mathrm{mix}}e^{-\left(\frac{r_\mathrm{mix}-r}{\sigma}\right)}
,\end{equation}
\noindent where $M_\mathrm{accr,mix}$ is the accreted matter in the region of mass $M_\mathrm{mix}$, $r_\mathrm{mix}$ is the radius at the bottom of the same region, and $\sigma$ is a fraction of the total radius of the WD that controls the steepness of the decay. We then iterated the value of $M_\mathrm{accr,mix}$ to obtain a total accreted mass $M_\mathrm{accr}=10^{16}$~g (a test mass, approximately the mass of Phobos). 

\begin{figure*}
\includegraphics[width=0.5\textwidth,clip=]{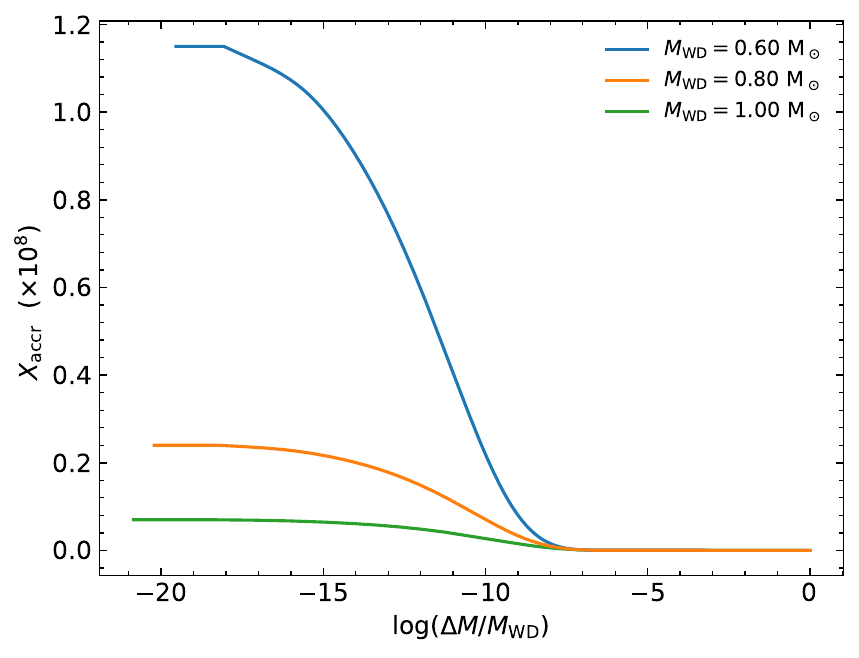}
\includegraphics[width=0.5\textwidth,clip=]{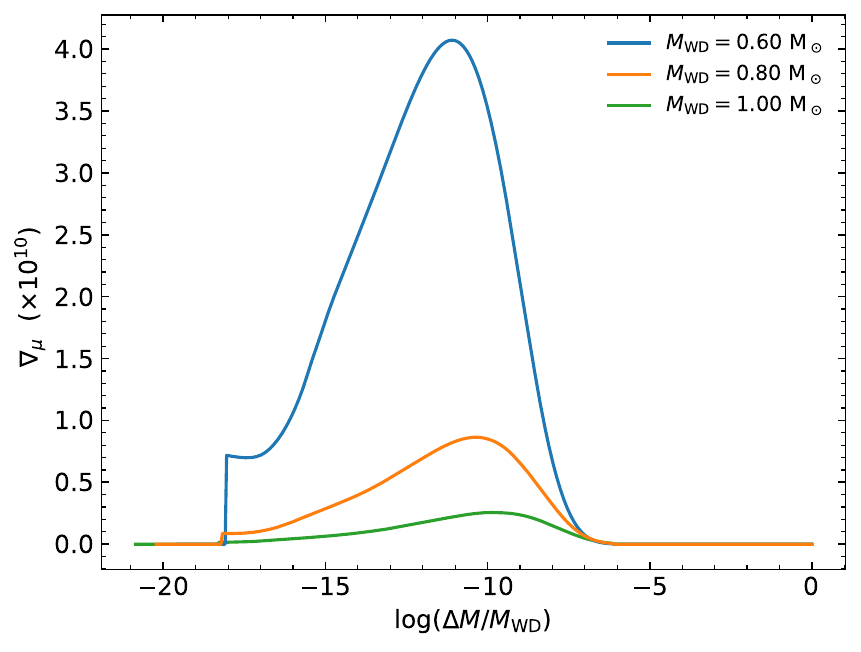}
\includegraphics[width=0.5\textwidth,clip=]{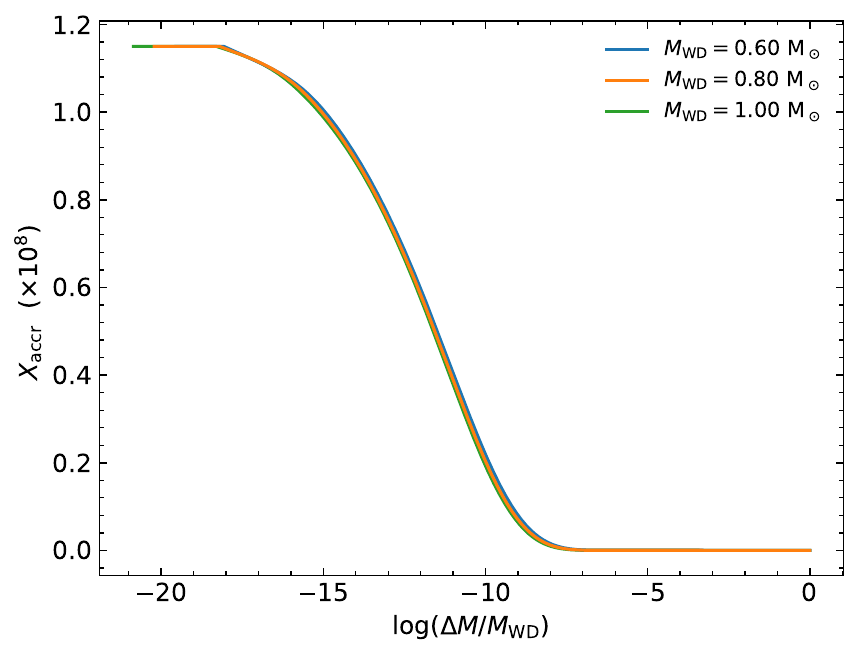}
\includegraphics[width=0.5\textwidth,clip=]{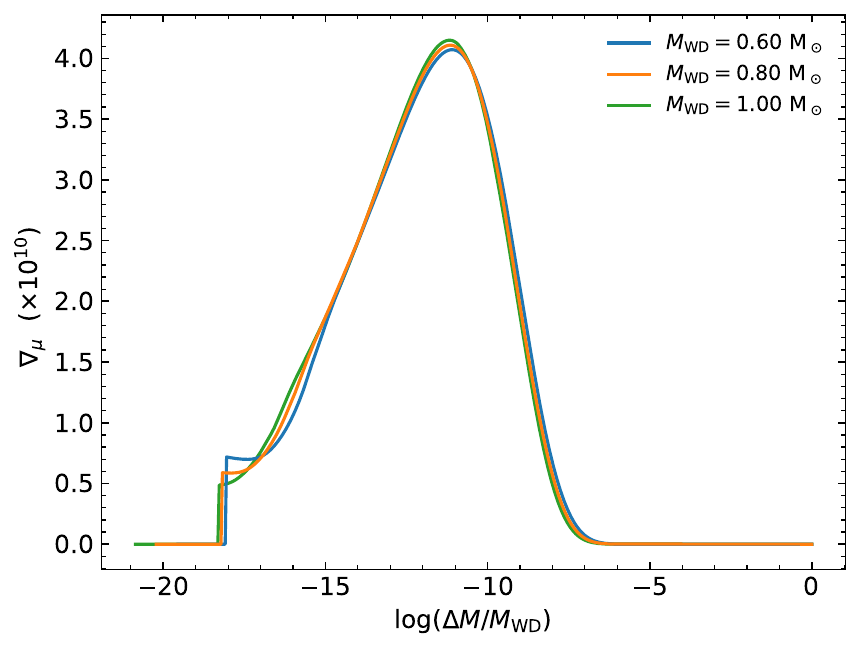}
\small
        \caption{Mass fraction profiles of accreted Mg (left panels) and mean molecular weight gradients' profiles (right panels) for the M06T25H-4, M08T25H-4, and M10T25H-4 models, as a function of the outer mass normalised to the total mass of the white dwarf model. The top panels correspond to case 1 (the same value for $\sigma$ for all models), the bottom panels to case 2 (a different value for $\sigma$  so that the $\mu$ gradients' profiles would be the same for all models). All cases were computed with the same total accreted mass ($10^{16}$g).}
\label{fig:xM}
\end{figure*}

Applying the same initial conditions for the accreted matter profiles for all models would introduce some mass-dependent effects that are not related to the efficiency of the dilution of the accreted matter. We followed two different scenarios to test the robustness of the results, including a case (case 2) where we introduced some fine-tuning to remove this dependency. In both cases, the properties of the accreted matter are the same for all models (i.e. the same accreted mass $M_\mathrm{accr}=10^{16}$~g and the same $M_\mathrm{mix}=10^{15}$~g). What differs is the exponential decay below the mixed zone (i.e. the value of $\sigma$).
\paragraph{Case 1:} The same exponential decay for all models with $\sigma$ equal to $0.25\%$ of the total radius of the WD. The computed profiles of the mass fraction of the accreted matter $X_\mathrm{accr}$ are shown in the top left panel of Fig.~\ref{fig:xM}. The corresponding mean molecular weight gradients are presented in the top right panel of the same figure. They are different for stars of different masses, which has a direct influence on the efficiency of thermohaline convection. 
\paragraph{Case 2:} In the second step, to eliminate this effect and evaluate the intrinsic effect of the WD mass, we adjusted the value of $\sigma$ to obtain the same mean molecular weight gradients in all models. The corresponding $\sigma$ are $0.25\%$, $0.177\%$, and $0.132\%$ of the total radius for the $0.6$, $0.8$, and $1.0$~M$_\odot$ models, respectively. The resulting mass fraction profiles and mean molecular weight gradients' profiles are presented in the bottom panels of Fig.~\ref{fig:xM}. The discontinuity seen around $\log(\Delta M/M_\mathrm{WD})=-18$ is the transition between the fully mixed region (corresponding to the region with the mass $M_\mathrm{mix}$) and the exponentially decaying region.

\subsection{The atomic diffusion coefficient}

The atomic diffusion coefficient was estimated in our models, following the approximations recommended by \cite{michaud15} for trace elements diffusing in a fully ionised pure hydrogen medium. This approximation is valid in the outer layers of all our models, as presented in the previous subsection. The expression for the atomic diffusion coefficient (Eq. 4.58 of \citealt{michaud15}) is then 

\begin{equation}
    D_{ip} \approx 1.947 \times 10^9 \frac{T^{5/2}}{n_p Z_i^2} \left(\frac{A_i + 1}{A_i}\right)^{1/2} \frac{1}{2 C_{ip}}
,\end{equation}
\noindent with 
\begin{equation}
    C_{ip} \approx \frac{1}{1.2}\ln(e^{1.2\ln\Lambda_{ij}} + 1)\end{equation}
and
\begin{equation}
    \Lambda_{ip} \approx 2.336\times10^3\frac{T^{3.2}}{Z_i n_p^{1/2}},
\end{equation}
\noindent where $T$ is the temperature, $n_p$ is the number density of the protons, and $Z_i$ and $A_i$ are the charge and mass of the diffusing atom, respectively. Assuming that the accreted matter has the same composition as Earth, the coefficient was estimated for magnesium ($A_i=24$, $Z_i=12$) since the mean mass of the particles of the bulk Earth composition is approximately $24.5$ \citep[estimated from][]{allegre95}.

\subsection{Thermohaline convection}

Thermohaline convection is a double diffusive instability that occurs in the presence of a stable temperature (T) gradient and an unstable mean molecular weight ($\mu$) gradient. When the destabilising effect of the $\mu$ gradient is larger than the stabilising effect of the T gradient, the medium is dynamically unstable. On the other hand, when the stabilising effect of the T gradient overcomes the destabilising effect of the $\mu$ gradient, the medium should be stable (the so-called Ledoux criterium). In this case, however, blobs of falling matter suffer from heat diffusion towards their surroundings on a much shorter timescale than particle diffusion. This induces another kind of instability, a double-diffusing one, namely thermohaline convection, which is also called fingering convection because it physically resembles falling fingers \citep[][and references therein]{vauclair04,charbonnel07,denissenkov10,theado09,garaud11}. These conditions are found in particular when heavy planetary matter is accreted onto the surface of WDs \cite[see e.g.][]{deal13,wachlin22,bauer18,bauer19,cresswell25}.

Precise studies of the effects of thermohaline convection in Earth's oceans began long ago \cite[e.g.][]{stern60}. The stellar case, which deals with very different physical situations, was only addressed twelve years later by \cite{ulrich72} (hereafter U72), who compared both cases and gave a first prescription for studying the efficiency of thermohaline mixing in stars. A few years later, \cite{kippenhahn80} (hereafter K80) gave a new prescription for the diffusion coefficient of thermohaline mixing, giving values about a hundred times lower than those of U72, 
\begin{equation}
    D_\mathrm{th}=\frac{C_t \kappa_T}{R_0},
\end{equation}
\noindent where $C_t = 12$ is a constant related to the so-called aspect ratio of the fingers, namely, the ratio of their length to their width \citep{kippenhahn80,theado12bis}.
\begin{equation}
    \kappa_T=\frac{4acT^3}{3\kappa C_p\rho^2}
\end{equation} 
\noindent is the thermal diffusivity (with $a$ the radiative density constant, $C_p$ the specific heat capacity at constant pressure, $\kappa$ the opacity, and $\rho$ the mass density) and $R_0=\frac{\nabla_\mathrm{ad}-\nabla}{|\nabla_\mu|}$ (where $\nabla=\partial\ln T /\partial \ln P$ is the temperature gradient, $P$ is the pressure, $\nabla_{ad}$ is the adiabatic temperature gradient, and $\nabla_\mu=\partial\ln \mu /\partial \ln P$ is the mean molecular weight gradient) is the so-called density ratio, namely the ratio of thermal to mean molecular weight gradients.
One of the important unknown parameters in these computations is the aspect ratio. Using 2D simulations, \cite{denissenkov10} showed that this aspect ratio is of the order of one, which is smaller than that assumed by U72. Later, \cite{traxler11} computed more precise mixing coefficients on the basis of 3D computations and confirmed this order of magnitude. Using this value in the computations leads to mixing coefficients close to those found by K80.
The condition needed for thermohaline mixing to develop inside stars is related to two important parameters: the ratio of atomic to thermal diffusivities $\tau$ and the so-called density ratio $R_0$. The thermohaline instability develops only if $R_0$ is in the following range:
\begin{equation}\label{eq:th_crit}
1<R_0<1/\tau
.\end{equation}
From their simulations, \cite{traxler11} showed that the values of the thermohaline mixing coefficient are close to that of K80 when the local value of $R_0$ is intermediate between the two limits. However, when it comes close to one of the two limits, the K80 value can overestimate the mixing coefficient. Performing computations using the K80 coefficient is safe, provided that the density ratio is far from the thermohaline limits.

More sophisticated expressions based on 3D simulations were proposed later on, including studies of other dynamical effects such as the 'stair cases', that is to say successions of thermohaline layers and dynamically convective layers that occur in some cases, and the possible interactions with shears or magnetic fields inside the stars \citep[see e.g.][]{Brown13,garaud19,fraser24}. For the first approach of this study, we used the simplest mixing coefficient given by K80, verifying that our computations remain in the range of $R_0$ for which this approximation is acceptable.

\section{Efficiency of thermohaline convection according to white dwarfs' properties}\label{sec:th}

\begin{figure*}[ht!]
\includegraphics[width=1.0\textwidth,clip=]{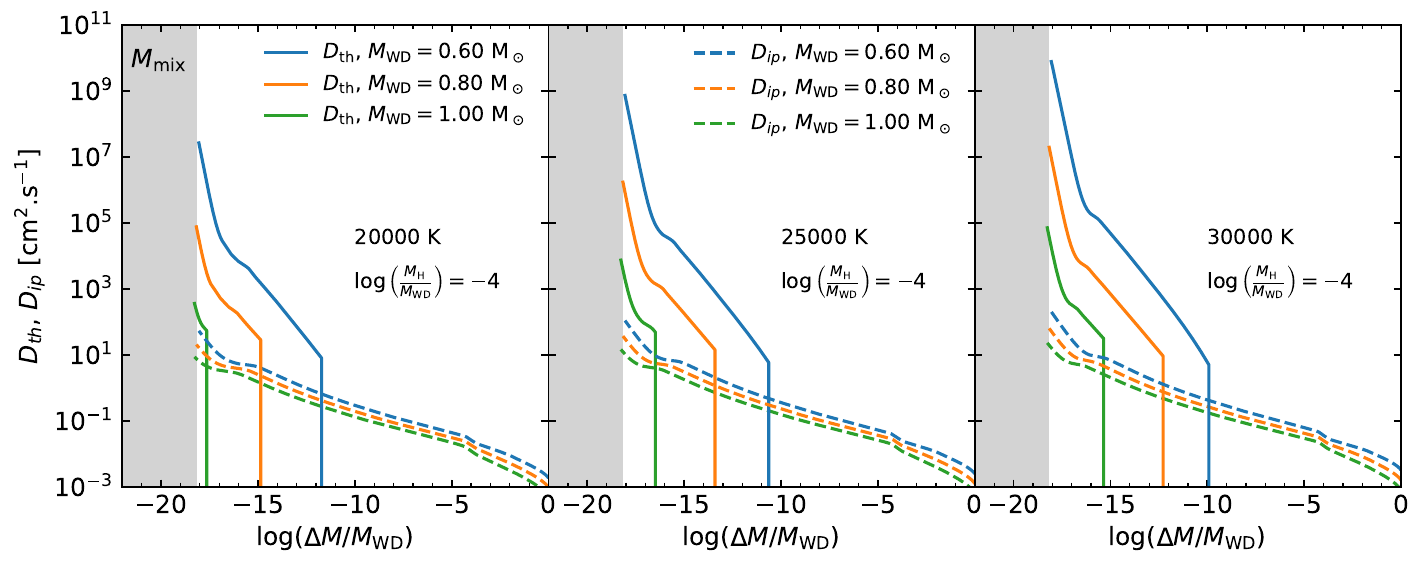}
\includegraphics[width=1.0\textwidth,clip=]{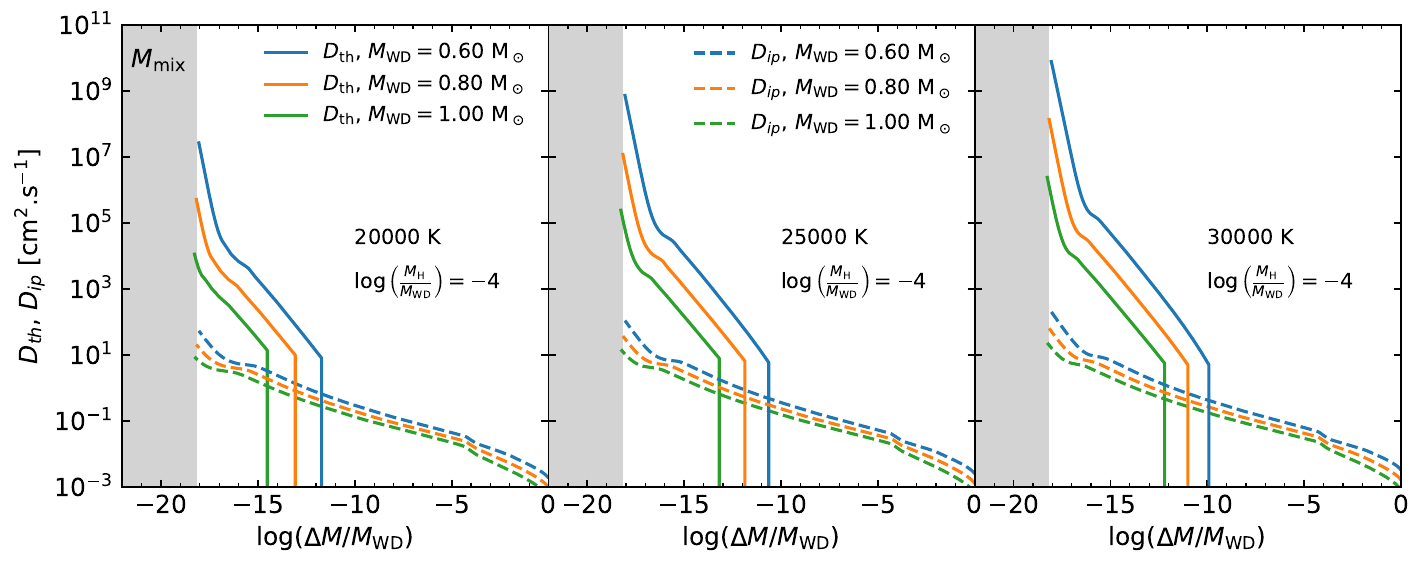}
        \caption{Diffusion coefficient for atomic diffusion (dashed lines) and thermohaline convection (solid lines) for different white dwarf masses and effective temperatures with $\log(M_\mathrm{H}/M_\mathrm{WD})=-4$. Cases 1 and 2 are  presented in the top and bottom panels, respectively. The grey areas represent the region of mass $M_\mathrm{mix}$ where the composition was initially fully mixed (the same region was plotted for each model for clarity).}
\label{fig:Dth-MTeff}
\end{figure*}

In this section, we compare the efficiency of thermohaline convection and atomic diffusion for the different models. The thermohaline and atomic diffusion coefficients are plotted in Fig.~\ref{fig:Dth-MTeff} as a function of the outer mass fraction for several cases. The effects of the WD masses and effective temperatures and that of the mass of the outer hydrogen layer were tested. The top panels display the results for case 1 (accreted profiles with the same value for $\sigma$ ) and the bottom panels display those for case 2 (accreted profiles with the same $mu$ gradients).

\subsection{Effect of mass and effective temperature}

In all cases, the large difference between the diffusion coefficients shows that when the stellar parameters allow thermohaline convection to occur, this thermohaline convection is much more efficient at transporting chemical elements than atomic diffusion, as is expected from the results of previous studies \citep[e.g.][and references therein]{vauclair04,deal13,wachlin22}. This is particularly important at the bottom of the fully mixed region of mass $M_\mathrm{mix}$ where accretion rates should be determined, which is equivalent to the bottom of convective envelopes for cooler WDs (see Sec.~\ref{sec:mdot}). Independently of the effective temperature, the efficiency of both processes decreases with increasing WD masses. This is mainly due to the density differences between the models. The higher density of more massive WDs directly comes from their mass-radius relation, which shows that the larger the mass, the smaller the radius \citep[see e.g.][]{althaus05,tremblay17,bedard17,sahu23,cang25}. This can be seen in the properties of the models in Table~\ref{tab:mod}. A higher density decreases the thermal diffusivity ($\kappa_T\propto 1/\rho^2$), therefore decreasing the efficiency of thermohaline convection. This is the same effect that occurs in DA and DB WDs. As the densities of the DB WDs are larger, the efficiency of thermohaline convection is lower \citep{deal13}. 
This means that for a given accreted mass (or accretion rate), the efficiency of dilution of matter decreases when the mass of the WDs increases, leading to a larger pollution signature at the surface. From Fig 2., we also notice that the higher the effective temperature, the more efficient the thermohaline convection transport. This is again due to the thermal diffusivity $\kappa_T\propto T^3$, which is larger at larger temperatures. The dilution of accreted matter should then be faster in hotter stars for a given accreted mass (or accretion rate).

For all models presented in Fig.~\ref{fig:Dth-MTeff}, the thermohaline mixed zone extends deeper as the mass of the WDs decreases. This effect is related to the instability criterium (Eq.~\ref{eq:th_crit}) that depends on $R_0$, and hence on the thermal diffusivity.

The results for cases 1 and 2 present the following differences:
\paragraph{Case 1:} The shape of the accreted matter profiles have the same properties for all models (especially the exponential decay in the radius). As the density is lower for smaller masses, the abundance variation at the surface is larger in the $0.6$M$_\odot$ models than in the $1.0$M$_\odot$ models, leading to a larger mean molecular weight gradient and a larger mixing by thermohaline convection. The difference with mass is then enhanced by this effect.

\paragraph{Case 2:} In this case, the $\mu$-gradient profiles are similar for all models as seen in the bottom right panel of Fig~\ref{fig:xM}. The differences in the efficiency of thermohaline convection according to mass are smaller but still present. In this case only the effect of the density of the models is seen, leading to about one order of magnitude differences in the diffusion coefficient of thermohaline convection. All other aspects are similar to case 1.

Altogether, we clearly show that the more massive the WDs, the less efficient the internal dilution processes. This should lead to larger pollution signatures at the surface. We note that the results are qualitatively the same if we use elements other than Mg such as Ca or Fe. We also tested that different configurations of the initial conditions (i.e. different values for $M_\mathrm{accr}$, $M_\mathrm{mix}$, and $\sigma$) do not affect the overall conclusions.

\subsection{Estimation of relative accretion rates}\label{sec:mdot}
\begin{figure}
\includegraphics[width=0.48\textwidth,clip=]{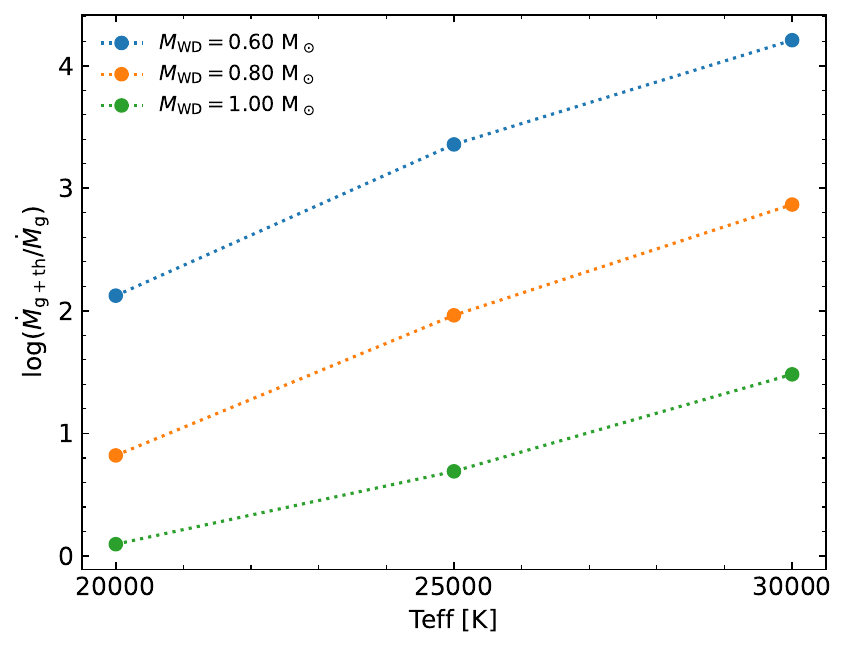}
        \caption{Ratio between accretion rates determined assuming gravitational settling and thermohaline convection as dilution processes ($\dot{M}_\mathrm{g+th}$) and accretion rates determined assuming gravitational settling only ($\dot{M}_\mathrm{g}$), for models of different masses and effective temperatures (listed in Tab.~\ref{tab:mod}). See the text for more details.}
\label{fig:Mdot}
\end{figure}

The observed pollution of WDs by heavy elements is the result of the balance between the fall of matter from outside and the dilution inside the star. It has been shown by \cite{deal13} and \cite{wachlin22} that in the case of element dilution dominated by thermohaline convection, equilibrium may be reached. A similar result is obtained when thermohaline convection is triggered by iron
accumulation induced by radiative acceleration in A-type stars \citep[see e.g.][]{theado09,zemskova14,deal16,hui-Bon-Hoa18}. Assuming that the abundance profiles presented in the bottom left panel of Fig.~\ref{fig:xM} are representative of such equilibria, the accretion rate leading to such profiles can be determined with

\begin{equation}
    \dot{M}=4\pi r^2X_\mathrm{accr}  \rho v,
\end{equation}
\noindent with
\begin{equation}
    v = D_\mathrm{ip} \frac{A_i m_p g}{k T} + D_\mathrm{th} \frac{1}{X_\mathrm{accr}}\frac{\partial X_\mathrm{accr}}{\partial r},
\end{equation}
\noindent where $g$ is the local gravity. All quantities were estimated at the bottom of the region of mass $M_\mathrm{mix}$ at the radius $r_\mathrm{mix}$.
Figure~\ref{fig:Mdot} shows the ratio of accretion rates determined with and without the contribution of thermohaline convection for the models with $M_H/M_\mathrm{WD}=10^{-4}$ presented in Table~\ref{tab:mod}. We can see in this figure that the ratio is always larger than one and may go up to four orders of magnitude. The accretion rates decrease when the mass and effective temperature of the WDs increase. This again shows the importance of the contribution of thermohaline convection in the context of accreting WDs. 
These results may be extrapolated to larger effective temperatures. However, this is not the case for lower effective temperatures because of the appearance of convective envelopes. The point at which the accretion rates are estimated would then be deeper inside the WD and the ratio would be slightly different. However, thermohaline convection would still lead to higher accretion rates by orders of magnitude \citep[see e.g.][]{deal13,wachlin22}.

\subsection{Effect of the depth of the outer hydrogen region}

Thermohaline convection develops in the presence of an inverse mean molecular weight gradient, and it is stopped in the case of stabilising gradients. This may occur at the bottom of the outer hydrogen zone in DA WDs, at the limit of the helium-rich region. We tested, even in the case of smaller hydrogen contents for models with $\log(M_H/M_\mathrm{WD})=-6$ and $-8$, that the hydrogen region is deep enough so that the helium transition region is not yet reached. However, this may not be the case for time-dependent accretion models for which the accreted matter could be trapped, in some cases, in the hydrogen zone. This will be treated in future work.

\section{Discussion}\label{sect:discu}

OR24 reported differences in the observed pollution of WD atmospheres by planetary matter accretion, according to their masses, for similar effective temperatures. They suggested that this trend could be due to different probabilities of planetary formation when the stars were on the main sequence. Massive stars would initially have fewer planetary systems than small mass stars.

This trend was confirmed by \cite{cunningham25}, who suggested various scenarios to explain this behaviour, in the framework of episodic accretion models. In both papers, however, the authors introduced element settling by atomic diffusion, but they did not discuss the impact of thermohaline convection, which is known to have important effects on DA WDs, such as the ones that they analysed.

\cite{buchan25} analysed the various hydrodynamical processes that could alter atomic diffusion results, including convective overshoot and thermohaline mixing. Using an approximate method,
they confirmed that thermohaline mixing decreases heavy element abundances in DA WDs, and that it must be taken into account in the computations of abundance evolution and accretion rates. They have not yet performed complete comparisons of the thermohaline effect according to  WD masses.

For the present study, we wanted to test whether thermohaline mixing could account for the abundance trend observed in DA WDs. Our results show that the element dilution inside these WDs is much larger when thermohaline convection is taken into account, but the effect decreases when the mass of the WD increases. As a consequence, the fact that massive WDs have suffered less matter accretion than small mass WDs seems to be confirmed, provided that atomic diffusion and thermohaline mixing are the only mixing processes responsible for the dilution of the accreted matter. In the following, we discuss several possibilities to explain this trend. 

\subsection{Caveats in our treatment of thermohaline convection}\label{sect:discu_accr}
A weak point of our computations is the uncertainty on the depth of penetration of the accreted particles and their initial mixing zone. When rapid particles fall onto stellar atmospheres, they rapidly thermalise in a very thin zone that is much smaller than the first layer of our models. Then thermohaline convection occurs immediately and mixes the elements downwards, leading to $\mu$ gradients varying with depth. We began our computations in an intermediate state for the initial conditions, with a small mixed zone at the surface and an exponentially decreasing abundance profile below. In Section~\ref{sec:th}, we have shown cases 1 and 2, which illustrate that modifying the abundance exponential decay profile may lead to a different efficiency of thermohaline convection, without changing the general trend. It is not impossible to manipulate these assumptions according to the stellar mass, so as to find a reverse trend in the element dilution, with more deficiency in massive WD than in lower-mass ones. This would, however, seem completely ad hoc and not satisfying. The only way to model these aspects correctly would be by solving the diffusion equation with time, in the presence of accretion.

\subsection{Other types of mixing inside the white dwarfs}
Other types of mixing processes may occur in WDs, and their variations with the stellar mass should be checked. Here we discuss rotational-induced mixing. In their analysis of the sample of 27 pulsating DA WDs (DAV) from Kepler and K2 data, \cite{hermes17} find a mean rotation period of 35 h with a mean standard deviation of 28 h for the WD mass interval between 0.51~M$_\odot$ and 0.73~M$_\odot$ and a much shorter rotation period of 1.13 h in the most massive WD of the sample with 0.9~M$_\odot$. Although they claim that it is an indication of a trend for more massive WDs to rotate faster than lower-mass WDs, it would require more measurements of the rotation periods of massive WDs to confirm that trend.  \cite{oliveira24} compiled the results of the rotation period deduced from asteroseismology data, adding the results from \cite{kawaler15} and \cite{kepler17} to those of \cite{hermes17}. They deduced a median rotation period of 24.3 h with a mean absolute deviation of 12.0 h for this sample of 63 pulsating WDs, and they do not confirm any trend according to the mass. Using TESS data, they studied the rotation period of a larger sample of WDs. Excluding magnetic WDs and WDs in close pairs for which the rotation may result from processes different from the simple angular momentum evolution during earlier phases of stellar evolution, they concentrated on a sample of 115 likely single WDs for which the observed variability is due to rotation. They find a median rotation period of 3.9 h. The difference with the median rotation period deduced from asteroseismology is due to the fact that from asteroseismology, one gets the internal rotation of the star, whereas the rotation periods deduced from TESS data are the rotation of the stellar surface induced by inhomogeneities at the surface. The distribution of the rotation periods as a function of the mass shows a large dispersion for masses between 0.4 and 0.6~M$_\odot$, from about 100 h to 0.1 h, while the dispersion is much smaller between 0.8 and 1.3~M$_\odot$, from about 2 h to 0.1 h. This is an indication that in general the more massive WDs rotate faster than the lower-mass WDs. In that case, rotation-induced mixing may be stronger for more massive WDs, leading to additional dilution that could compensate for the different efficiencies of thermohaline convection or even lead to stronger dilution. Time-dependent accretion models that include the full treatment of angular momentum and chemical element transport would be required to test this hypothesis.

\subsection{Masses and history of the white dwarf progenitors}
Going back to the most probable results, that the accretion signatures are larger for WDs of smaller masses, we shall discuss the connexion to the main sequence progenitors of these WDs. The link between the pollution of WDs and the formation of planets is subject to many uncertainties, especially when assessing planet formation occurrence according to the initial mass of a star. Relating the mass of a WD to that of its main-sequence progenitor is very challenging. There are many approximations for mass loss rates throughout the evolution, and especially for the latest phases of evolution, that make predictions from stellar evolution models difficult. Moreover, $30$ to $50\%$ of the more massive WDs are the result of binary mergers, making the link with the progenitor even more uncertain \citep[see][and references therein]{temmink20}. OR24 discussed this aspect in detail and did not find any merger signature in their sample (looking at magnetism, rotation, and kinematics). However, the difficulty in determining the formation mass of WDs may therefore affect the conclusion on the planetary system occurrences according to the mass of the progenitor on the main sequence.

\section{Conclusion}\label{sect:conclu}
White dwarf stars polluted by the accretion of planetary material onto their atmospheres provide good tools for inferring the pristine chemical composition and physical conditions of planetary systems at their births. However, one has to be careful when deriving the composition of the accreted matter from the spectroscopically determined abundances. The observed elements are the result of the balance between the accretion rate and the rate of dilution inside the star. Atomic diffusion is not the only diluting process. In some cases, thermohaline mixing is of much greater importance and has to be treated correctly; otherwise, the obtained results are unreliable. Computations show that thermohaline effects are much weaker in He-rich WDs than in H-rich ones, so the conclusions obtained in the literature by studying only He-rich ones are correct. The results presented by \cite{ouldrouis24} rely on H-rich WDs, and we have shown that neglecting thermohaline convection for these stars leads to underestimating the dilution process and hence the derived accretion rates. However, according to our computations, their conclusions regarding the rate of formation of planetary systems according to the stellar masses on the main sequence may remain valid when thermohaline convection is taken into account, under the condition that the progenitors of more massive WDs are actually more massive stars. In any case, a more sophisticated (time-dependent) modelling of the accretion and dilution processes is needed to draw stronger conclusions.

\begin{acknowledgements}
This work was supported by CNES, focused on PLATO, \textit{Kepler}, and TESS. 
\end{acknowledgements}  

\bibliographystyle{aa} 
\bibliography{main.bib} 

@BOOK{michaud15,
       author = {{Michaud}, Georges and {Alecian}, Georges and {Richer}, Jacques},
        title = "{Atomic Diffusion in Stars}",
         year = 2015,
          doi = {10.1007/978-3-319-19854-5},
       adsurl = {https://ui.adsabs.harvard.edu/abs/2015ads..book.....M},
    publisher = {Springer}
}

@ARTICLE{theado12bis,
   author = {{Th{\'e}ado}, S. and {Vauclair}, S.},
    title = "{Metal-rich Accretion and Thermohaline Instabilities in Exoplanet-host Stars: Consequences on the Light Elements Abundances}",
  journal = {\apj},
archivePrefix = "arXiv",
   eprint = {1109.4238},
 primaryClass = "astro-ph.SR",
     year = 2012,
    month = jan,
   volume = 744,
      eid = {123},
    pages = {123},
      doi = {10.1088/0004-637X/744/2/123},
   adsurl = {http://adsabs.harvard.edu/abs/2012ApJ...744..123T}
}

@ARTICLE{vauclair04,
   author = {{Vauclair}, S.},
    title = "{Metallic Fingers and Metallicity Excess in Exoplanets' Host Stars: The Accretion Hypothesis Revisited}",
  journal = {\apj},
   eprint = {arXiv:astro-ph/0309790},
     year = 2004,
    month = apr,
   volume = 605,
    pages = {874-879},
      doi = {10.1086/382668},
   adsurl = {http://adsabs.harvard.edu/abs/2004ApJ...605..874V}
}

@ARTICLE{zuckerman10,
   author = {{Zuckerman}, B. and {Melis}, C. and {Klein}, B. and {Koester}, D. and 
	{Jura}, M.},
    title = "{Ancient Planetary Systems are Orbiting a Large Fraction of White Dwarf Stars}",
  journal = {\apj},
archivePrefix = "arXiv",
   eprint = {1007.2252},
 primaryClass = "astro-ph.SR",
     year = 2010,
    month = oct,
   volume = 722,
    pages = {725-736},
      doi = {10.1088/0004-637X/722/1/725},
   adsurl = {http://adsabs.harvard.edu/abs/2010ApJ...722..725Z}
}

@ARTICLE{allegre95,
   author = {{All{\`e}gre}, C.~J. and {Poirier}, J.-P. and {Humler}, E. and 
	{Hofmann}, A.~W.},
    title = "{The chemical composition of the Earth}",
  journal = {Earth and Planetary Science Letters},
     year = 1995,
    month = sep,
   volume = 134,
    pages = {515-526},
      doi = {10.1016/0012-821X(95)00123-T},
   adsurl = {http://adsabs.harvard.edu/abs/1995E%26PSL.134..515A}
}

@ARTICLE{Brown13,
   author = {{Brown}, J.~M. and {Garaud}, P. and {Stellmach}, S.},
    title = "{Chemical Transport and Spontaneous Layer Formation in Fingering Convection in Astrophysics}",
  journal = {ApJ},
archivePrefix = "arXiv",
   eprint = {1212.1688},
 primaryClass = "astro-ph.SR",
     year = 2013,
    month = may,
   volume = 768,
      eid = {34},
    pages = {34}
}

@ARTICLE{zemskova14,
   author = {{Zemskova}, V. and {Garaud}, P. and {Deal}, M. and {Vauclair}, S.
	},
    title = "{Fingering Convection Induced by Atomic Diffusion in Stars: 3D Numerical Computations and Applications to Stellar Models}",
  journal = {\apj},
archivePrefix = "arXiv",
   eprint = {1407.1437},
 primaryClass = "astro-ph.SR",
     year = 2014,
    month = nov,
   volume = 795,
      eid = {118},
    pages = {118},
      doi = {10.1088/0004-637X/795/2/118},
   adsurl = {http://adsabs.harvard.edu/abs/2014ApJ...795..118Z}
}

@ARTICLE{garaud11,
   author = {{Garaud}, P.},
    title = "{What Happened to the Other Mohicans? The Case for a Primordial Origin to the Planet-Metallicity Connection}",
  journal = {\apjl},
     year = 2011,
    month = feb,
   volume = 728,
      eid = {L30},
    pages = {L30},
      doi = {10.1088/2041-8205/728/2/L30},
   adsurl = {http://adsabs.harvard.edu/abs/2011ApJ...728L..30G}
}

@ARTICLE{deal13,
   author = {{Deal}, M. and {Deheuvels}, S. and {Vauclair}, G. and {Vauclair}, S. and 
	{Wachlin}, F.~C.},
    title = "{Accretion from debris disks onto white dwarfs. Fingering (thermohaline) instability and derived accretion rates}",
  journal = {\aap},
archivePrefix = "arXiv",
   eprint = {1308.5406},
 primaryClass = "astro-ph.SR",
     year = 2013,
    month = sep,
   volume = 557,
      eid = {L12},
    pages = {L12},
      doi = {10.1051/0004-6361/201322206},
   adsurl = {http://adsabs.harvard.edu/abs/2013A%26A...557L..12D}
}

@ARTICLE{charbonnel07,
   author = {{Charbonnel}, C. and {Zahn}, J.-P.},
    title = "{Inhibition of thermohaline mixing by a magnetic field in Ap star descendants: implications for the Galactic evolution of $^{3}$He}",
  journal = {\aap},
archivePrefix = "arXiv",
   eprint = {0711.3395},
     year = 2007,
    month = dec,
   volume = 476,
    pages = {L29-L32},
      doi = {10.1051/0004-6361:20078740},
   adsurl = {http://adsabs.harvard.edu/abs/2007A%26A...476L..29C}
}

@ARTICLE{ulrich72,
   author = {{Ulrich}, R.~K.},
    title = "{Thermohaline Convection in Stellar Interiors.}",
  journal = {\apj},
     year = 1972,
    month = feb,
   volume = 172,
    pages = {165},
      doi = {10.1086/151336},
   adsurl = {http://adsabs.harvard.edu/abs/1972ApJ...172..165U}
}

@ARTICLE{kippenhahn80,
   author = {{Kippenhahn}, R. and {Ruschenplatt}, G. and {Thomas}, H.-C.},
    title = "{The time scale of thermohaline mixing in stars}",
  journal = {\aap},
     year = 1980,
    month = nov,
   volume = 91,
    pages = {175-180},
   adsurl = {http://adsabs.harvard.edu/abs/1980A%26A....91..175K}
}

@ARTICLE{denissenkov10,
   author = {{Denissenkov}, P.~A.},
    title = "{Numerical Simulations of Thermohaline Convection: Implications for Extra-mixing in Low-mass RGB Stars}",
  journal = {\apj},
archivePrefix = "arXiv",
   eprint = {1006.5481},
 primaryClass = "astro-ph.SR",
     year = 2010,
    month = nov,
   volume = 723,
    pages = {563-579},
      doi = {10.1088/0004-637X/723/1/563},
   adsurl = {http://adsabs.harvard.edu/abs/2010ApJ...723..563D}
}

@ARTICLE{traxler11,
   author = {{Traxler}, A. and {Garaud}, P. and {Stellmach}, S.},
    title = "{Numerically Determined Transport Laws for Fingering (''Thermohaline'') Convection in Astrophysics}",
  journal = {\apjl},
archivePrefix = "arXiv",
   eprint = {1011.3461},
 primaryClass = "astro-ph.SR",
     year = 2011,
    month = feb,
   volume = 728,
      eid = {L29},
    pages = {L29},
      doi = {10.1088/2041-8205/728/2/L29},
   adsurl = {http://adsabs.harvard.edu/abs/2011ApJ...728L..29T}
}

@ARTICLE{deal16,
   author = {{Deal}, M. and {Richard}, O. and {Vauclair}, S.},
    title = "{Hydrodynamical instabilities induced by atomic diffusion in A stars and their consequences}",
  journal = {\aap},
archivePrefix = "arXiv",
   eprint = {1604.01241},
 primaryClass = "astro-ph.SR",
     year = 2016,
    month = apr,
   volume = 589,
      eid = {A140},
    pages = {A140},
      doi = {10.1051/0004-6361/201628180},
   adsurl = {http://adsabs.harvard.edu/abs/2016A%26A...589A.140D}
}

@ARTICLE{theado09,
   author = {{Th{\'e}ado}, S. and {Vauclair}, S. and {Alecian}, G. and {LeBlanc}, F.
	},
    title = "{Influence of Thermohaline Convection on Diffusion-Induced Iron Accumulation in a Stars}",
  journal = {\apj},
archivePrefix = "arXiv",
   eprint = {0908.1534},
 primaryClass = "astro-ph.SR",
     year = 2009,
    month = oct,
   volume = 704,
    pages = {1262-1273},
      doi = {10.1088/0004-637X/704/2/1262},
   adsurl = {http://adsabs.harvard.edu/abs/2009ApJ...704.1262T}
}

@ARTICLE{hui-Bon-Hoa18,
   author = {{Hui-Bon-Hoa}, A. and {Vauclair}, S.},
    title = "{Role of atomic diffusion in the opacity enhancement inside B-type stars}",
  journal = {\aap},
     year = 2018,
    month = feb,
   volume = 610,
      eid = {L15},
    pages = {L15},
      doi = {10.1051/0004-6361/201832706},
   adsurl = {http://adsabs.harvard.edu/abs/2018A%26A...610L..15H}
}

@ARTICLE{ouldrouis24,
       author = {{Ould Rouis}, Lou Baya and {Hermes}, J.~J. and {G{\"a}nsicke}, Boris T. and {Sahu}, Snehalata and {Koester}, Detlev and {Tremblay}, P. -E. and {Veras}, Dimitri and {Farihi}, Jay and {Heintz}, Tyler M. and {Gentile Fusillo}, Nicola Pietro and {Redfield}, Seth},
        title = "{Constraints on Remnant Planetary Systems as a Function of Main-sequence Mass with HST/COS}",
      journal = {\apj},
         year = 2024,
        month = dec,
       volume = {976},
       number = {2},
          eid = {156},
        pages = {156},
          doi = {10.3847/1538-4357/ad86bb},
archivePrefix = {arXiv},
       eprint = {2410.06335},
 primaryClass = {astro-ph.SR},
       adsurl = {https://ui.adsabs.harvard.edu/abs/2024ApJ...976..156O}
}

@ARTICLE{romero25,
       author = {{Romero}, Alejandra D. and {Kepler}, S.~O. and {Oliveira da Rosa}, Gabriela and {Hermes}, J.~J.},
        title = "{Thirty-two New Bright ZZ Ceti Stars from TESS: Adding Cycles 4 and 5}",
      journal = {\apj},
         year = 2025,
        month = may,
       volume = {984},
       number = {2},
          eid = {112},
        pages = {112},
          doi = {10.3847/1538-4357/adc113},
archivePrefix = {arXiv},
       eprint = {2407.07260},
 primaryClass = {astro-ph.SR},
       adsurl = {https://ui.adsabs.harvard.edu/abs/2025ApJ...984..112R}
}

@ARTICLE{wachlin22,
       author = {{Wachlin}, F.~C. and {Vauclair}, G. and {Vauclair}, S. and {Althaus}, L.~G.},
        title = "{New simulations of accreting DA white dwarfs: Inferring accretion rates from the surface contamination}",
      journal = {\aap},
         year = 2022,
        month = apr,
       volume = {660},
          eid = {A30},
        pages = {A30},
          doi = {10.1051/0004-6361/202142289},
archivePrefix = {arXiv},
       eprint = {2109.11370},
 primaryClass = {astro-ph.SR},
       adsurl = {https://ui.adsabs.harvard.edu/abs/2022A&A...660A..30W}
}

@ARTICLE{cresswell25,
       author = {{Cresswell}, Imogen G. and {Fraser}, Adrian E. and {Bauer}, Evan B. and {Anders}, Evan H. and {Brown}, Benjamin P.},
        title = "{3D Simulations Demonstrate Propagating Thermohaline Convection for Polluted White Dwarfs}",
      journal = {\apjl},
         year = 2025,
        month = jun,
       volume = {986},
       number = {1},
          eid = {L10},
        pages = {L10},
          doi = {10.3847/2041-8213/addbd5},
archivePrefix = {arXiv},
       eprint = {2503.20885},
 primaryClass = {astro-ph.SR},
       adsurl = {https://ui.adsabs.harvard.edu/abs/2025ApJ...986L..10C}
}

@ARTICLE{fraser24,
       author = {{Fraser}, Adrian E. and {Reifenstein}, Sam A. and {Garaud}, Pascale},
        title = "{Magnetized Fingering Convection in Stars}",
      journal = {\apj},
         year = 2024,
        month = apr,
       volume = {964},
       number = {2},
          eid = {184},
        pages = {184},
          doi = {10.3847/1538-4357/ad26fe},
archivePrefix = {arXiv},
       eprint = {2302.11610},
 primaryClass = {astro-ph.SR},
       adsurl = {https://ui.adsabs.harvard.edu/abs/2024ApJ...964..184F}
}

@ARTICLE{garaud19,
       author = {{Garaud}, P. and {Kumar}, A. and {Sridhar}, J.},
        title = "{The Interaction between Shear and Fingering (Thermohaline) Convection}",
      journal = {\apj},
         year = 2019,
        month = jul,
       volume = {879},
       number = {1},
          eid = {60},
        pages = {60},
          doi = {10.3847/1538-4357/ab232f},
archivePrefix = {arXiv},
       eprint = {1905.07636},
 primaryClass = {astro-ph.SR},
       adsurl = {https://ui.adsabs.harvard.edu/abs/2019ApJ...879...60G}
}

@ARTICLE{stern60,
       author = {{Stern}, Melvin E.},
        title = "{The ``Salt-Fountain'' and Thermohaline Convection}",
      journal = {Tellus},
         year = 1960,
        month = may,
       volume = {12},
       number = {2},
        pages = {172-175},
          doi = {10.1111/j.2153-3490.1960.tb01295.x10.3402/tellusa.v12i2.9378},
       adsurl = {https://ui.adsabs.harvard.edu/abs/1960Tell...12..172S}
}

@ARTICLE{gentille-fusillo21,
       author = {{Gentile Fusillo}, N.~P. and {Manser}, C.~J. and {G{\"a}nsicke}, Boris T. and {Toloza}, O. and {Koester}, D. and {Dennihy}, E. and {Brown}, W.~R. and {Farihi}, J. and {Hollands}, M.~A. and {Hoskin}, M.~J. and {Izquierdo}, P. and {Kinnear}, T. and {Marsh}, T.~R. and {Santamar{\'\i}a-Miranda}, A. and {Pala}, A.~F. and {Redfield}, S. and {Rodr{\'\i}guez-Gil}, P. and {Schreiber}, M.~R. and {Veras}, Dimitri and {Wilson}, D.~J.},
        title = "{White dwarfs with planetary remnants in the era of Gaia - I. Six emission line systems}",
      journal = {\mnras},
         year = 2021,
        month = jun,
       volume = {504},
       number = {2},
        pages = {2707-2726},
          doi = {10.1093/mnras/stab992},
archivePrefix = {arXiv},
       eprint = {2010.13807},
 primaryClass = {astro-ph.SR},
       adsurl = {https://ui.adsabs.harvard.edu/abs/2021MNRAS.504.2707G}
}

@ARTICLE{bauer19,
       author = {{Bauer}, Evan B. and {Bildsten}, Lars},
        title = "{Polluted White Dwarfs: Mixing Regions and Diffusion Timescales}",
      journal = {\apj},
         year = 2019,
        month = feb,
       volume = {872},
       number = {1},
          eid = {96},
        pages = {96},
          doi = {10.3847/1538-4357/ab0028},
archivePrefix = {arXiv},
       eprint = {1812.09602},
 primaryClass = {astro-ph.SR},
       adsurl = {https://ui.adsabs.harvard.edu/abs/2019ApJ...872...96B}
}

@ARTICLE{harrison21,
       author = {{Harrison}, John H.~D. and {Bonsor}, Amy and {Kama}, Mihkel and {Buchan}, Andrew M. and {Blouin}, Simon and {Koester}, Detlev},
        title = "{Bayesian constraints on the origin and geology of exoplanetary material using a population of externally polluted white dwarfs}",
      journal = {\mnras},
         year = 2021,
        month = jun,
       volume = {504},
       number = {2},
        pages = {2853-2867},
          doi = {10.1093/mnras/stab736},
archivePrefix = {arXiv},
       eprint = {2103.05713},
 primaryClass = {astro-ph.EP},
       adsurl = {https://ui.adsabs.harvard.edu/abs/2021MNRAS.504.2853H}
}

@ARTICLE{bhattacharjee25,
       author = {{Bhattacharjee}, Soumyadeep},
        title = "{Thick Disks Around White Dwarfs Viewed ``Edge-off'': Effects on Transit Properties and Infrared Excess}",
      journal = {\pasp},
         year = 2025,
        month = nov,
       volume = {137},
       number = {11},
          eid = {114204},
        pages = {114204},
          doi = {10.1088/1538-3873/ae0fe3},
archivePrefix = {arXiv},
       eprint = {2507.20594},
 primaryClass = {astro-ph.SR},
       adsurl = {https://ui.adsabs.harvard.edu/abs/2025PASP..137k4204B}
}

@ARTICLE{rogers25,
       author = {{Rogers}, Laura K. and {Bonsor}, Amy and {Le Bourdais}, {\'E}rika and {Xu}, Siyi and {Su}, Kate Y.~L. and {Richards}, Benjamin and {Buchan}, Andrew and {Ballering}, Nicholas P. and {Brouwers}, Marc and {Dufour}, Patrick and {Kissler-Patig}, Markus and {Melis}, Carl and {Zuckerman}, Ben},
        title = "{Silicate mineralogy and bulk composition of exoplanetary material in polluted white dwarfs}",
      journal = {\mnras},
         year = 2025,
        month = jul,
          doi = {10.1093/mnras/staf1221},
archivePrefix = {arXiv},
       eprint = {2507.16777},
 primaryClass = {astro-ph.EP},
       adsurl = {https://ui.adsabs.harvard.edu/abs/2025MNRAS.tmp.1174R}
}

@ARTICLE{zuckerman87,
       author = {{Zuckerman}, B. and {Becklin}, E.~E.},
        title = "{Excess infrared radiation from a white dwarf{\textemdash}an orbiting brown dwarf?}",
      journal = {\nat},
         year = 1987,
        month = nov,
       volume = {330},
       number = {6144},
        pages = {138-140},
          doi = {10.1038/330138a0},
       adsurl = {https://ui.adsabs.harvard.edu/abs/1987Natur.330..138Z}
}

@ARTICLE{vanderburg15,
       author = {{Vanderburg}, Andrew and {Johnson}, John Asher and {Rappaport}, Saul and {Bieryla}, Allyson and {Irwin}, Jonathan and {Lewis}, John Arban and {Kipping}, David and {Brown}, Warren R. and {Dufour}, Patrick and {Ciardi}, David R. and {Angus}, Ruth and {Schaefer}, Laura and {Latham}, David W. and {Charbonneau}, David and {Beichman}, Charles and {Eastman}, Jason and {McCrady}, Nate and {Wittenmyer}, Robert A. and {Wright}, Jason T.},
        title = "{A disintegrating minor planet transiting a white dwarf}",
      journal = {\nat},
         year = 2015,
        month = oct,
       volume = {526},
       number = {7574},
        pages = {546-549},
          doi = {10.1038/nature15527},
archivePrefix = {arXiv},
       eprint = {1510.06387},
 primaryClass = {astro-ph.EP},
       adsurl = {https://ui.adsabs.harvard.edu/abs/2015Natur.526..546V}
}

@ARTICLE{vanderbosch20,
       author = {{Vanderbosch}, Z. and {Hermes}, J.~J. and {Dennihy}, E. and {Dunlap}, B.~H. and {Izquierdo}, P. and {Tremblay}, P. -E. and {Cho}, P.~B. and {G{\"a}nsicke}, B.~T. and {Toloza}, O. and {Bell}, K.~J. and {Montgomery}, M.~H. and {Winget}, D.~E.},
        title = "{A White Dwarf with Transiting Circumstellar Material Far outside the Roche Limit}",
      journal = {\apj},
         year = 2020,
        month = jul,
       volume = {897},
       number = {2},
          eid = {171},
        pages = {171},
          doi = {10.3847/1538-4357/ab9649},
archivePrefix = {arXiv},
       eprint = {1908.09839},
 primaryClass = {astro-ph.SR},
       adsurl = {https://ui.adsabs.harvard.edu/abs/2020ApJ...897..171V}
}

@ARTICLE{vanderbosch21,
       author = {{Vanderbosch}, Zachary P. and {Rappaport}, Saul and {Guidry}, Joseph A. and {Gary}, Bruce L. and {Blouin}, Simon and {Kaye}, Thomas G. and {Weinberger}, Alycia J. and {Melis}, Carl and {Klein}, Beth L. and {Zuckerman}, B. and {Vanderburg}, Andrew and {Hermes}, J.~J. and {Hegedus}, Ryan J. and {Burleigh}, Matthew. R. and {Sefako}, Ramotholo and {Worters}, Hannah L. and {Heintz}, Tyler M.},
        title = "{Recurring Planetary Debris Transits and Circumstellar Gas around White Dwarf ZTF J0328-1219}",
      journal = {\apj},
         year = 2021,
        month = aug,
       volume = {917},
       number = {1},
          eid = {41},
        pages = {41},
          doi = {10.3847/1538-4357/ac0822},
archivePrefix = {arXiv},
       eprint = {2106.02659},
 primaryClass = {astro-ph.EP},
       adsurl = {https://ui.adsabs.harvard.edu/abs/2021ApJ...917...41V}
}

@ARTICLE{farihi08,
       author = {{Farihi}, J. and {Zuckerman}, B. and {Becklin}, E.~E.},
        title = "{Spitzer IRAC Observations of White Dwarfs. I. Warm Dust at Metal-Rich Degenerates}",
      journal = {\apj},
         year = 2008,
        month = feb,
       volume = {674},
       number = {1},
        pages = {431-446},
          doi = {10.1086/521715},
archivePrefix = {arXiv},
       eprint = {0710.0907},
 primaryClass = {astro-ph},
       adsurl = {https://ui.adsabs.harvard.edu/abs/2008ApJ...674..431F}
}

@ARTICLE{guidry24,
       author = {{Guidry}, Joseph A. and {Hermes}, J.~J. and {De}, Kishalay and {Ould Rouis}, Lou Baya and {Ewing}, Brison B. and {Kaiser}, B.~C.},
        title = "{Using 3.4 {\ensuremath{\mu}}m Variability toward White Dwarfs as a Signpost of Remnant Planetary Systems}",
      journal = {\apj},
         year = 2024,
        month = sep,
       volume = {972},
       number = {1},
          eid = {126},
        pages = {126},
          doi = {10.3847/1538-4357/ad5be7},
archivePrefix = {arXiv},
       eprint = {2406.18646},
 primaryClass = {astro-ph.SR},
       adsurl = {https://ui.adsabs.harvard.edu/abs/2024ApJ...972..126G}
}

@ARTICLE{cang25,
       author = {{Cang}, Tianqi and {Zhang}, Jiayi and {Fu}, Jian-Ning and {Zhao}, He and {Zong}, Weikai},
        title = "{Seismic test of the mass-radius relationship of hydrogen-atmospheric white dwarf stars}",
      journal = {Journal of Astrophysics and Astronomy},
         year = 2025,
        month = oct,
       volume = {46},
       number = {2},
          eid = {81},
        pages = {81},
          doi = {10.1007/s12036-025-10106-3},
archivePrefix = {arXiv},
       eprint = {2508.03184},
 primaryClass = {astro-ph.SR},
       adsurl = {https://ui.adsabs.harvard.edu/abs/2025JApA...46...81C}
}

@ARTICLE{althaus05,
       author = {{Althaus}, L.~G. and {Garc{\'\i}a-Berro}, E. and {Isern}, J. and {C{\'o}rsico}, A.~H.},
        title = "{Mass-radius relations for massive white dwarf stars}",
      journal = {\aap},
         year = 2005,
        month = oct,
       volume = {441},
       number = {2},
        pages = {689-694},
          doi = {10.1051/0004-6361:20052996},
archivePrefix = {arXiv},
       eprint = {astro-ph/0507559},
 primaryClass = {astro-ph},
       adsurl = {https://ui.adsabs.harvard.edu/abs/2005A&A...441..689A}
}

@ARTICLE{tremblay17,
       author = {{Tremblay}, P. -E. and {Gentile-Fusillo}, N. and {Raddi}, R. and {Jordan}, S. and {Besson}, C. and {G{\"a}nsicke}, B.~T. and {Parsons}, S.~G. and {Koester}, D. and {Marsh}, T. and {Bohlin}, R. and {Kalirai}, J. and {Deustua}, S.},
        title = "{The Gaia DR1 mass-radius relation for white dwarfs}",
      journal = {\mnras},
         year = 2017,
        month = mar,
       volume = {465},
       number = {3},
        pages = {2849-2861},
          doi = {10.1093/mnras/stw2854},
archivePrefix = {arXiv},
       eprint = {1611.00629},
 primaryClass = {astro-ph.SR},
       adsurl = {https://ui.adsabs.harvard.edu/abs/2017MNRAS.465.2849T}
}

@ARTICLE{bedard17,
       author = {{B{\'e}dard}, A. and {Bergeron}, P. and {Fontaine}, G.},
        title = "{Measurements of Physical Parameters of White Dwarfs: A Test of the Mass-Radius Relation}",
      journal = {\apj},
         year = 2017,
        month = oct,
       volume = {848},
       number = {1},
          eid = {11},
        pages = {11},
          doi = {10.3847/1538-4357/aa8bb6},
archivePrefix = {arXiv},
       eprint = {1709.02324},
 primaryClass = {astro-ph.SR},
       adsurl = {https://ui.adsabs.harvard.edu/abs/2017ApJ...848...11B}
}

@ARTICLE{sahu23,
       author = {{Sahu}, Snehalata and {G{\"a}nsicke}, Boris T. and {Tremblay}, Pier-Emmanuel and {Koester}, Detlev and {Hermes}, J.~J. and {Wilson}, David J. and {Toloza}, Odette and {Hoskin}, Matthew J. and {Farihi}, Jay and {Manser}, Christopher J. and {Redfield}, Seth},
        title = "{An HST COS ultraviolet spectroscopic survey of 311 DA white dwarfs - I. Fundamental parameters and comparative studies}",
      journal = {\mnras},
         year = 2023,
        month = dec,
       volume = {526},
       number = {4},
        pages = {5800-5823},
          doi = {10.1093/mnras/stad2663},
archivePrefix = {arXiv},
       eprint = {2309.00239},
 primaryClass = {astro-ph.SR},
       adsurl = {https://ui.adsabs.harvard.edu/abs/2023MNRAS.526.5800S}
}

@INPROCEEDINGS{kawaler15,
       author = {{Kawaler}, Steven D.},
        title = "{Rotation of White Dwarf Stars}",
    booktitle = {19th European Workshop on White Dwarfs},
         year = 2015,
       editor = {{Dufour}, P. and {Bergeron}, P. and {Fontaine}, G.},
       series = {Astronomical Society of the Pacific Conference Series},
       volume = {493},
        month = jun,
        pages = {65},
          doi = {10.48550/arXiv.1410.6934},
archivePrefix = {arXiv},
       eprint = {1410.6934},
 primaryClass = {astro-ph.SR},
       adsurl = {https://ui.adsabs.harvard.edu/abs/2015ASPC..493...65K}
}

@INPROCEEDINGS{kepler17,
       author = {{Kepler}, S.~O. and {Romero}, Alejandra D.},
        title = "{Pulsating white dwarfs}",
    booktitle = {European Physical Journal Web of Conferences},
         year = 2017,
       series = {European Physical Journal Web of Conferences},
       volume = {152},
        month = sep,
          eid = {01011},
        pages = {01011},
          doi = {10.1051/epjconf/201715201011},
archivePrefix = {arXiv},
       eprint = {1706.07020},
 primaryClass = {astro-ph.SR},
       adsurl = {https://ui.adsabs.harvard.edu/abs/2017EPJWC.15201011K}
}

@ARTICLE{hermes17,
       author = {{Hermes}, J.~J. and {G{\"a}nsicke}, B.~T. and {Kawaler}, Steven D. and {Greiss}, S. and {Tremblay}, P. -E. and {Gentile Fusillo}, N.~P. and {Raddi}, R. and {Fanale}, S.~M. and {Bell}, Keaton J. and {Dennihy}, E. and {Fuchs}, J.~T. and {Dunlap}, B.~H. and {Clemens}, J.~C. and {Montgomery}, M.~H. and {Winget}, D.~E. and {Chote}, P. and {Marsh}, T.~R. and {Redfield}, S.},
        title = "{White Dwarf Rotation as a Function of Mass and a Dichotomy of Mode Line Widths: Kepler Observations of 27 Pulsating DA White Dwarfs through K2 Campaign 8}",
      journal = {\apjs},
         year = 2017,
        month = oct,
       volume = {232},
       number = {2},
          eid = {23},
        pages = {23},
          doi = {10.3847/1538-4365/aa8bb5},
archivePrefix = {arXiv},
       eprint = {1709.07004},
 primaryClass = {astro-ph.SR},
       adsurl = {https://ui.adsabs.harvard.edu/abs/2017ApJS..232...23H}
}

@ARTICLE{oliveira24,
       author = {{Oliveira da Rosa}, Gabriela and {Kepler}, S.~O. and {Soethe}, L.~T.~T. and {Romero}, Alejandra D. and {Bell}, Keaton J.},
        title = "{Photometric White Dwarf Rotation}",
      journal = {\apj},
         year = 2024,
        month = oct,
       volume = {974},
       number = {2},
          eid = {314},
        pages = {314},
          doi = {10.3847/1538-4357/ad6987},
archivePrefix = {arXiv},
       eprint = {2407.05214},
 primaryClass = {astro-ph.SR},
       adsurl = {https://ui.adsabs.harvard.edu/abs/2024ApJ...974..314O}
}

@ARTICLE{bedard22,
       author = {{B{\'e}dard}, A. and {Brassard}, P. and {Bergeron}, P. and {Blouin}, S.},
        title = "{On the Spectral Evolution of Hot White Dwarf Stars. II. Time-dependent Simulations of Element Transport in Evolving White Dwarfs with STELUM}",
      journal = {\apj},
         year = 2022,
        month = mar,
       volume = {927},
       number = {1},
          eid = {128},
        pages = {128},
          doi = {10.3847/1538-4357/ac4497},
archivePrefix = {arXiv},
       eprint = {2112.09989},
 primaryClass = {astro-ph.SR},
       adsurl = {https://ui.adsabs.harvard.edu/abs/2022ApJ...927..128B}
}

@ARTICLE{giammichele16,
       author = {{Giammichele}, N. and {Fontaine}, G. and {Brassard}, P. and {Charpinet}, S.},
        title = "{A New Analysis of the Two Classical ZZ Ceti White Dwarfs GD 165 and Ross 548. II. Seismic Modeling}",
      journal = {\apjs},
         year = 2016,
        month = mar,
       volume = {223},
       number = {1},
          eid = {10},
        pages = {10},
          doi = {10.3847/0067-0049/223/1/10},
       adsurl = {https://ui.adsabs.harvard.edu/abs/2016ApJS..223...10G}
}

@ARTICLE{giammichele17,
       author = {{Giammichele}, N. and {Charpinet}, S. and {Fontaine}, G. and {Brassard}, P.},
        title = "{Toward High-precision Seismic Studies of White Dwarf Stars: Parametrization of the Core and Tests of Accuracy}",
      journal = {\apj},
         year = 2017,
        month = jan,
       volume = {834},
       number = {2},
          eid = {136},
        pages = {136},
          doi = {10.3847/1538-4357/834/2/136},
archivePrefix = {arXiv},
       eprint = {1610.06036},
 primaryClass = {astro-ph.SR},
       adsurl = {https://ui.adsabs.harvard.edu/abs/2017ApJ...834..136G}
}

@ARTICLE{salaris10,
       author = {{Salaris}, M. and {Cassisi}, S. and {Pietrinferni}, A. and {Kowalski}, P.~M. and {Isern}, J.},
        title = "{A Large Stellar Evolution Database for Population Synthesis Studies. VI. White Dwarf Cooling Sequences}",
      journal = {\apj},
         year = 2010,
        month = jun,
       volume = {716},
       number = {2},
        pages = {1241-1251},
          doi = {10.1088/0004-637X/716/2/1241},
archivePrefix = {arXiv},
       eprint = {1005.1791},
 primaryClass = {astro-ph.SR},
       adsurl = {https://ui.adsabs.harvard.edu/abs/2010ApJ...716.1241S}
}

@ARTICLE{bauer18,
       author = {{Bauer}, Evan B. and {Bildsten}, Lars},
        title = "{Increases to Inferred Rates of Planetesimal Accretion due to Thermohaline Mixing in Metal-accreting White Dwarfs}",
      journal = {\apjl},
         year = 2018,
        month = jun,
       volume = {859},
       number = {2},
          eid = {L19},
        pages = {L19},
          doi = {10.3847/2041-8213/aac492},
archivePrefix = {arXiv},
       eprint = {1805.05425},
 primaryClass = {astro-ph.SR},
       adsurl = {https://ui.adsabs.harvard.edu/abs/2018ApJ...859L..19B}
}

@ARTICLE{temmink20,
       author = {{Temmink}, K.~D. and {Toonen}, S. and {Zapartas}, E. and {Justham}, S. and {G{\"a}nsicke}, B.~T.},
        title = "{Looks can be deceiving. Underestimating the age of single white dwarfs due to binary mergers}",
      journal = {\aap},
         year = 2020,
        month = apr,
       volume = {636},
          eid = {A31},
        pages = {A31},
          doi = {10.1051/0004-6361/201936889},
archivePrefix = {arXiv},
       eprint = {1910.05335},
 primaryClass = {astro-ph.SR},
       adsurl = {https://ui.adsabs.harvard.edu/abs/2020A&A...636A..31T}
}

@ARTICLE{buchan25,
       author = {{Buchan}, Andrew M. and {Tremblay}, Pier-Emmanuel and {B{\'e}dard}, Antoine and {Bauer}, Evan B. and {Cunningham}, Tim},
        title = "{Exogeological inferences from white dwarf pollutants: the impact of stellar physics}",
      journal = {\mnras},
         year = 2025,
        month = dec,
       volume = {544},
       number = {2},
        pages = {2098-2119},
          doi = {10.1093/mnras/staf1832},
archivePrefix = {arXiv},
       eprint = {2510.17985},
 primaryClass = {astro-ph.EP},
       adsurl = {https://ui.adsabs.harvard.edu/abs/2025MNRAS.544.2098B}
}

@ARTICLE{cunningham25,
       author = {{Cunningham}, Tim and {Tremblay}, Pier-Emmanuel and {O'Brien}, Mairi and {Bauer}, Evan B. and {Hollands}, Mark A. and {Koester}, Detlev and {Kenyon}, Scott J. and {Charbonneau}, David and {Veras}, Dimitri and {Yusaf}, Muhammad Furqaan},
        title = "{The dearth of high-mass hydrogen-atmosphere metal-polluted white dwarfs within 40 pc}",
      journal = {\mnras},
         year = 2025,
        month = may,
       volume = {539},
       number = {3},
        pages = {2021-2038},
          doi = {10.1093/mnras/staf428},
archivePrefix = {arXiv},
       eprint = {2503.09734},
 primaryClass = {astro-ph.SR},
       adsurl = {https://ui.adsabs.harvard.edu/abs/2025MNRAS.539.2021C}
}
 

\end{document}